\documentclass[%
 reprint,
 amsmath,amssymb,
 aps,
 prl,
 twocolumn,
 groupedaddress,
 superscriptaddress,
]{revtex4-2}

\usepackage{graphicx}
\usepackage{dcolumn}
\usepackage{bm}
\usepackage{tikz}
\usepackage{amssymb}
\usepackage{amsmath}
\usepackage{multirow}
\usepackage{natbib}

\graphicspath{{./FigMainPaper/},{./AllFig/}}
\newcommand\Rey{\mbox{\textit{Re}}}

\begin{document}

\title{The phase transition to turbulence in spatially extended shear flows}

\author{Lukasz Klotz}
\affiliation{ Institute of Science and Technology Austria, Am Campus 1, 3400 Klosterneuburg, Austria}
\affiliation{Institute of Aeronautics and Applied Mechanics, Warsaw University of Technology,  Nowowiejska 24, 00-665 Warsaw, Poland}
\author{Gr\'{e}goire Lemoult}%
 \affiliation{ CNRS, UMR 6294, Laboratoire Onde et Milieux Complexes (LOMC)53, Normandie Université, UniHavre, rue de Prony, Le Havre Cedex 76058, France}

\author{Kerstin Avila}%
\affiliation{ University of Bremen, Faculty of Production Engineering, Badgasteiner Strasse 1, 28359 Bremen, Germany }
\affiliation{Leibniz Institute for Materials Engineering IWT, Badgasteiner Strasse 3, 28359 Bremen, Germany}  
  
\author{Bj\"orn Hof}
\thanks{bhof@ist.ac.at}
\affiliation{ Institute of Science and Technology Austria, Am Campus 1, 3400 Klosterneuburg, Austria}
\email{bhof@ist.ac.at}
\date{\today}

\begin{abstract}

Directed percolation(DP) has recently emerged as a possible solution to the century old puzzle surrounding the transition to turbulence. Multiple model studies reported DP exponents, however experimental evidence is limited since the largest possible observation times are orders of magnitude shorter than the flows’ characteristic time scales. An exception is cylindrical Couette flow where the limit is not temporal but rather the realizable system size. We present experiments in a Couette setup of unprecedented azimuthal and axial aspect ratios. Approaching the critical point to within less than 0.1\% we determine five critical exponents, all of which are in excellent agreement with the 2+1D DP-universality class. The complex dynamics encountered at the onset of turbulence can hence be fully rationalized within the framework of statistical mechanics.

\end{abstract}

\maketitle

With increasing velocity basic fluid flows undergo a sudden transition from ordered laminar motion to turbulence. Even for the geometrically simplest cases such as pipe, Couette and channel flow standard stability theory can neither explain the nature of the transition nor predict its occurrence. To make matters worse, transitional flows erratically switch between perfectly laminar and highly chaotic phases resulting in fluctuation levels that are substantially larger than those encountered in fully turbulent flows at much higher velocities. Despite this apparent complexity, statistical mechanics offers a surprisingly simple explanation for this phenomenon. Macroscopic features of the dynamics bear close resemblance to the universality class of directed percolation \cite{hinrichsen_non-equilibrium_2000}, the probably most basic phase transition outside of thermal equilibrium.

Directed percolation encompasses a large number of phenomena, including simple epidemic models, fluid flow through porous media and forest fires \cite{hinrichsen_non-equilibrium_2000}. A common feature of all these problems is the competition between an active (e.g. infected) and an absorbing (e.g. susceptible) state governed by a control parameter. Assuming the perspective of turbulence and its eventual decay encountered for decreasing flow speeds, it becomes apparent that the resulting laminar flow qualifies as an absorbing state. Being stable to infinitesimal perturbations, laminar motion persists and the active state, turbulence, cannot arise spontaneously. Based on this analogy Pomeau \cite{pomeau_front_1986} suggested that the propagation of laminar turbulent fronts may be determined by the critical exponents of directed percolation. This proposition was initially studied for simpler models \cite{kaneko_spatiotemporal_1985,chate_transition_1987,chate_spatio-temporal_1988,rolf_directed_1998} exhibiting spatio-temporal chaos, however it could not be confirmed for the onset of turbulence in actual fluid flows. In contrast to the expected continuous transition, initial experiments as well as direct numerical simulations of Couette flow reported a discontinuous transition \cite{bottin_discontinuous_1998,duguet_formation_2010}. A modified analogy to directed percolation has emerged more recently and resulted from studies of the large scale localized patches that turbulence tends to organize in, close to the critical point. In pipe flow these so called ‘puffs’ have a finite lifetime and their decay is memoryless \cite{hof_finite_2006,hof_repeller_2008}. Prior to decay puffs can seed new puffs by nearest neighbor interactions, a process called puff splitting \cite{moxey_distinct_2010}, which is equally memoryless \cite{avila_onset_2011}. These studies suggest that an individual percolation site is not a turbulent eddy but a macroscopic puff and hence a structure that is more than an order of magnitude larger than sites were previously expected to be. A further difference to the original DP analogy is that this transition does not account for the eventual emergence of fully turbulent flow, which only results from a distinct transition at higher Re \cite{barkley_rise_2015,barkley_theoretical_2016}. Puffs consist of a short turbulent upstream part (5D in length) and an elongated quiescent leading edge (20D) across which the profile recovers from a plug to a more parabolic shape. This profile recovery sets the minimum puff spacing \citep{hof_science_2010, samanta_experimental_2011}. Consequently the analogous state to a fully occupied lattice in DP is a dense pattern of distinct puffs, not fully turbulent flow. The dynamics of the process are governed by the decay and splitting rates of puffs and in the proximity of the critical point on average a puff travels $10^7$ diameters downstream before either event occurs \cite{avila_onset_2011}. While various numerical studies determined directed percolation exponents for models of varying complexity(including direct numerical simulations)\cite{barkley_simplifying_2011,sipos_directed_2011,shih_ecological_2016,chantry_universal_2017,barkley_theoretical_2016,manneville_transitional_2020,takeda_intermittency_2020}, measurements in actual fluid flows have to resolve both, the large spatial and temporal scales. 

So far this has only been possible in an experimental setting where like in pipe flow the dynamics of turbulent sites is confined to a single spatial  dimension \cite{lemoult_directed_2016}. The circular Couette geometry chosen in that study and the resulting periodic boundary conditions allowed to observe puff like turbulent structures for sufficiently large times (order of $10^6$ advective units, note that at the critical point of Couette flow the decay and splitting times are about two decades smaller than those in pipes \cite{shi_scale_2013}). In this case indeed a continuous transition was reported with exponents closely matching those of directed percolation in 1+1 dimensions. The purpose of the present study is to establish if the DP analogy observed for flows extended in one dimension, carries over to planar flows where turbulence arranges in stripe patterns that proliferate in two dimensions. While for the Waleffe model the onset of sustained turbulent stripes has been shown to closely agree with the 2+1D DP universality class \cite{chantry_universal_2017}, experimental evidence is missing. 

\begin{figure}[!htp]
\begin{center}
\includegraphics*[trim = 0cm -0.5cm 0 0cm,scale=0.5]{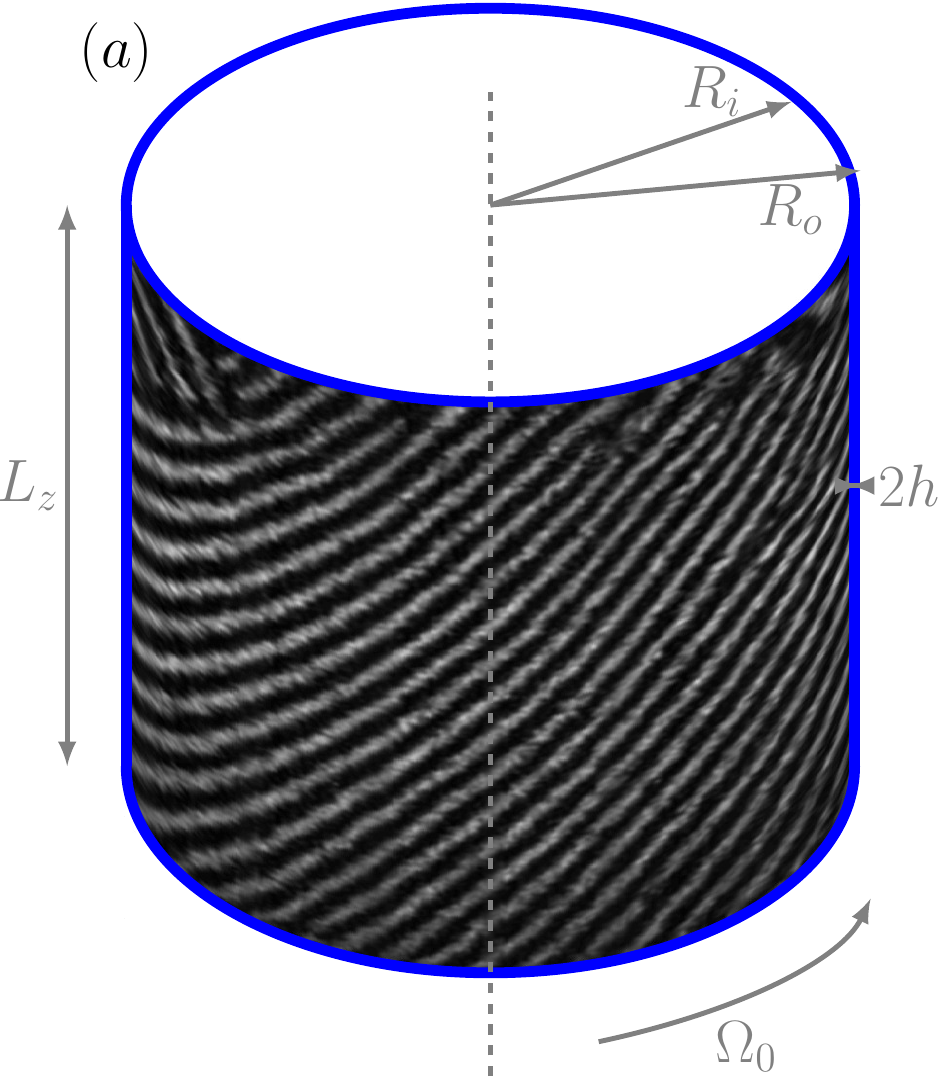}
\includegraphics[trim = 1.9cm 1.8cm 0cm 0.5cm, scale=0.33]{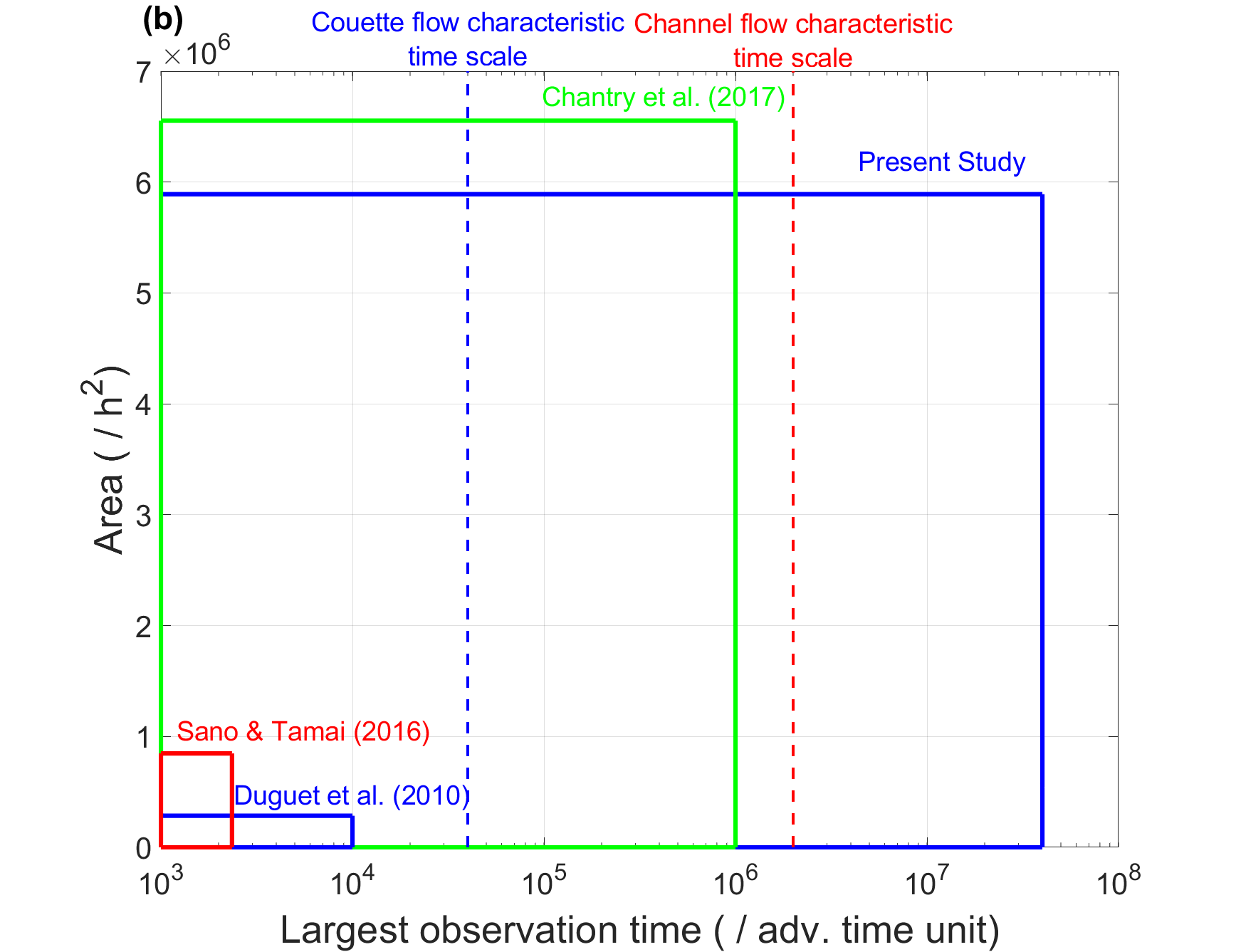}

\end{center}
\caption{Schematic of experimental apparatus (top); typical test section sizes and time scales covered in previous studies when compared to the present experiment (bottom). Note that the observation time is defined as the maximum time over which the evolution of a flow pattern can be investigated in a co moving frame. It hence does not necessarily coincide with (and often is far shorter than) the recording time.} 
\label{fig:expSetUp}
\end{figure}

At odds with the large scales (see Fig.~\ref{fig:expSetUp}b for a comparison of the temporal and spatial scales) required for the investigation of the transition type described above, a recent experimental study of channel flow \cite{sano_universal_2016} reported a fundamentally different DP analogy. In that case the inlet was forced to be fully turbulent and the flow was investigated in a comparably small experimental domain (see vertical dimension of red square in Fig.~\ref{fig:expSetUp}b) and for short observation times (see horizontal extend of red square compared to the three orders of magnitude larger stripe splitting times \citep{gome_statistical_2020}, red dashed line). However in this case the eventual flow never settled to the large scale stripe patterns that have been reported in other studies of channel flow \citep{tsukahara_dns_2005,xiong_turbulent_2015,shimizu_bifurcations_2019,PhD_2019_Paranjape}. Moreover recent numerical and experimental studies of channel flow \cite{xiong_turbulent_2015,tao_extended_2018,shimizu_bifurcations_2019,PhD_2019_Paranjape} could not confirm the phase transition point suggested by \cite{sano_universal_2016}. Instead in these more recent studies turbulent stripes were observed to persist to significantly lower Reynolds numbers, suggesting that in \cite{sano_universal_2016} flows may have not reached the eventual statistical steady state and that the observed dynamics were transient, which would also explain the difference in the structures observed. It is also noteworthy that  \cite{sano_universal_2016} obtained nonzero turbulent fractions below what they suggested as a critical point. While the authors attributed the none vanishing turbulent fraction to finite size effects, a finite system size classically has the opposite effect and causes turbulent fractions to already reach zero (i.e. a discontinuity) slightly above critical, provided that observation times are sufficient. 

In the present investigation of cylindrical Couette flow the periodic boundary conditions ensure that flow patterns can be investigated for arbitrarily long times. To ensure that the results are not affected by finite size effects \cite{avila_high-precision_2013,avila_second-order_2021}, a very small gap size ($2h$) had to be chosen. In contrast to \cite{lemoult_directed_2016} our system is not only extended in the azimuthal direction but equally in the spanwise (or axial) direction and hence flow patterns are expected to be two dimensional and consist of stripes and not of puffs. Given that the minimum spacing between turbulent stripes in Couette flow is approximately 70 h, aspect ratios in excess of thousand h are required to accommodate flow patterns comprising sufficient stripe numbers (i.e. a sufficient number of sites). 

\begin{figure*}[!htp]
	\centering
	\includegraphics[width= 0.95\textwidth]{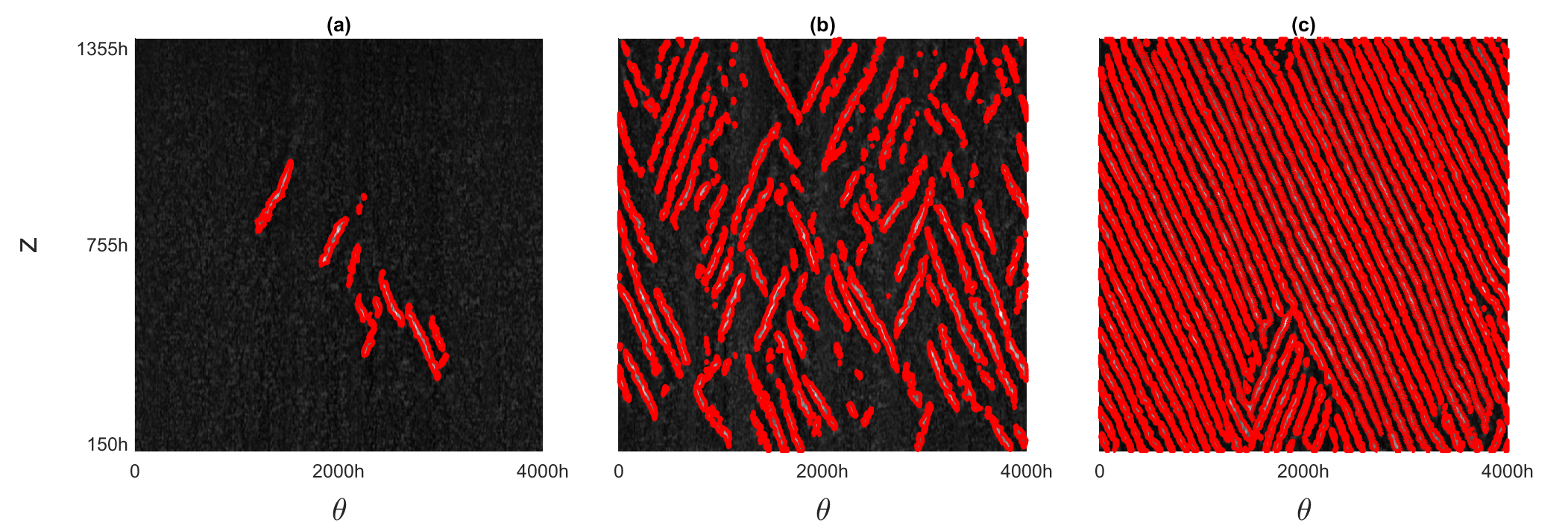}
	\caption{ Turbulent stripes in the axial-azimuthal cross section. Flow visualizations for $\Rey=331$ (a), $\Rey=333$ (b), and $\Rey=349$ (c). Turbulent phase is marked by red contours. The laminar regions (black) expand as the Reynolds number is decreased and stripes become sparse.}	
	\label{fig:visu}
\end{figure*} 
\begin{figure*}[!htp]
	\centering
	\includegraphics[width= 0.95\textwidth]{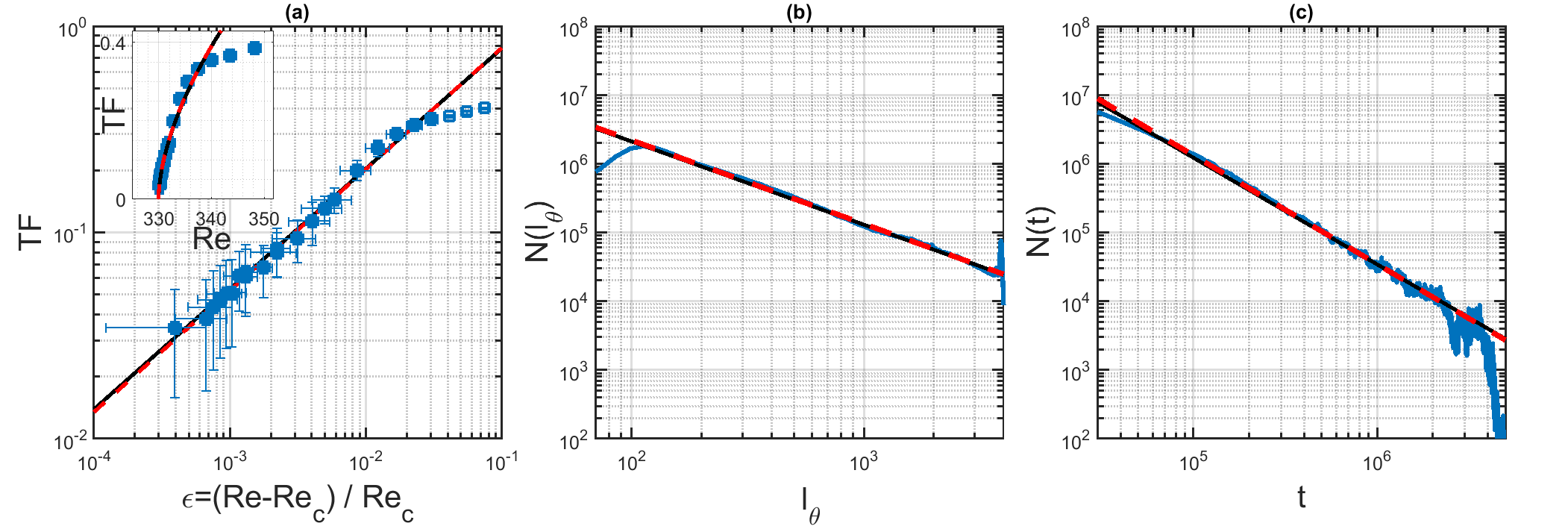}
	\caption{ Order parameter and critical exponents. (a) power-law dependence of turbulent fraction TF (blue squares) on the distance from the critical Reynolds number $\epsilon=(\Rey-\Rey_{c})/\Rey_{c}$ shown in log-log (linear scale in inset). The distributions of the azimuthal (streamwise) sizes of the laminar gaps $N(l_{\theta})$ and the laminar time intervals $N(\tau)$ close to the critical point ($\Rey=331$) are illustrated by blue curves in panel (b) and (c), respectively. In each panel the red dashed lines illustrate the best fit (panel (a): $TF\sim \epsilon^{\beta}$ with $\beta=0.59$; panel (b): $N(\l_{\theta})\sim l^{-\mu_{\bot}}$ with $\mu_{\bot}=1.22 \pm 0.02$; panel (c): $N(\tau)\sim \tau^{-\mu_{\parallel}}$ with $\mu_{\parallel}=1.58 \pm 0.03$) and black solid lines correspond to theoretical prediction for (2+1)DP universality class ($\beta^{DP}=0.58,\,\mu_{\bot}^{DP}=1.21,\,\mu_{\parallel}^{DP}=1.55$).}
	\label{fig:TfMu}
\end{figure*} 
\begin{figure*}[t]
	\centering
	\includegraphics[trim={0 -0.6cm 0 -1cm},width= 0.95\textwidth]{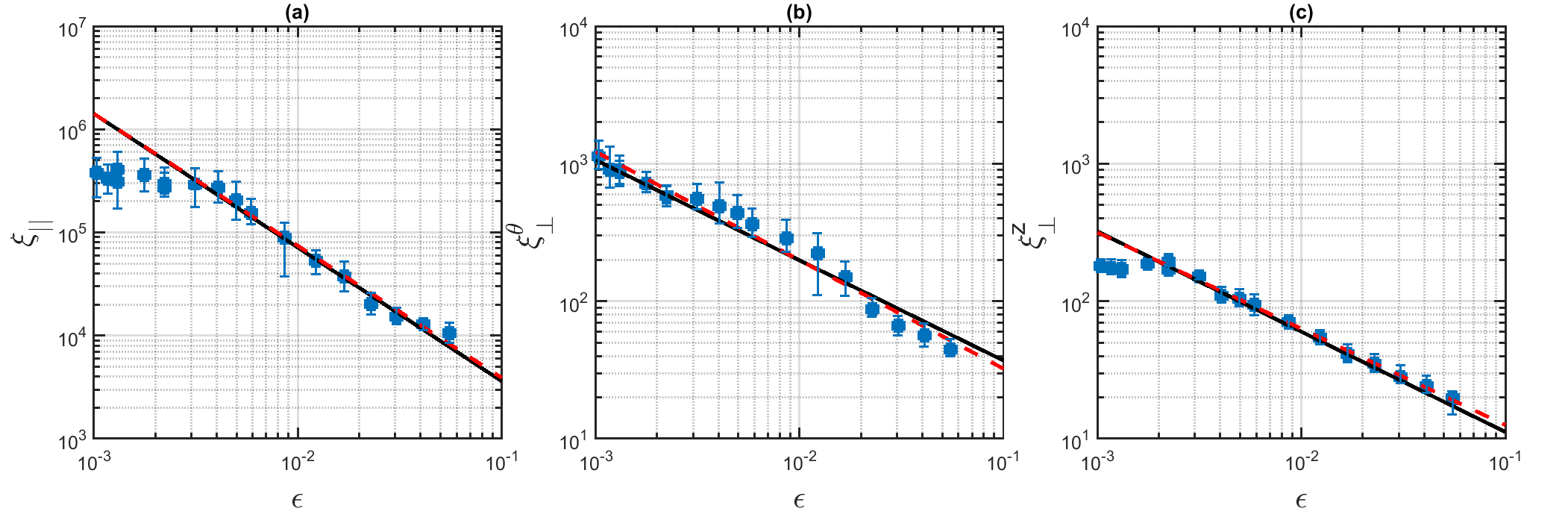}
	\caption{ Correlation exponents at the phase transition. Dependence of correlation time (panel (a): $\xi_{\parallel}$)) and correlation lengths (panel (b): $\xi_{\bot}^{\theta}$; panel (c): $\xi_{\bot}^{z}$ ) on control parameter $\epsilon$. In each panel the red dashed line corresponds to the best fit (panel (a): $\xi_{\parallel} \sim \epsilon^{-\nu_{\parallel}}$ with $\nu_{\parallel}=1.28$; panel (b): $\xi_{\bot}^{\theta} \sim \epsilon^{-\nu_{\bot}^{\theta}}$ with $\nu_{\bot}^{\theta}=0.79$; panel (c): $\xi_{\bot}^{z} \sim \epsilon^{-\nu_{\bot}^{z}}$ with $\nu_{\bot}^{z}=0.70$) and black solid lines correspond to the theoretical predictions for the (2+1)DP universality class ($\nu_{\parallel}=1.296, \nu_{\bot}^{DP}=0.729$).}
	\label{fig:nuCoeff}
\end{figure*} 

The experiment consists of a pair of concentric, high precision cylinders; an outer glass cylinder (with an inner diameter of $150$ mm) and an inner (hollow) stainless steel cylinder (with an outer diameter of $149.52$mm), resulting in a gap size of $2h=240{\mu}$m (Fig.~\ref{fig:expSetUp}). Given the azimuthal circumference $C=471.225$mm and the cylinder length of $H = 180$mm the azimuthal (C/h) and axial (H/h) aspect ratios are $3927$ and $1500$ respectively. While the outer cylinder is held in place by air bearings (constraining it in the azimuthal and the axial direction), the inner one is held by a hollow shaft that allows to circulate temperature controlled fluid through the cylinder core and hence to temperature control the experiment (with a precision of $0.02K$). The gap between the two cylinders is filled with ethanol ($99$\%) and in addition a small amount (mass concentration $C_m \approx 5\cdot 10^{-4}$) of micron sized aluminum flakes (Eckart PCR1100) is added to the fluid to visualize the flow patterns. The latter were recorded using two line cameras (Basler spL 4096) (see also \cite{lemoult_directed_2016} and supplemental materials for the azimuthal reconstruction from a time series). The flow is driven by the rotation of the outer cylinder with angular velocities ranging from 1000 RPM to 1300 RPM. The gap uniformity during outer cylinder rotation (i.e. relative variation around the mean value) has been measured to be better than $\pm 6\mu$m and the eccentricity between the inner and the outer cylinder is less than $5\mu$m, as described in detail in the supplemental materials. The control parameter, the Reynolds number ($\Rey=\Omega_{o} R_{o} h\,/\,2\nu$, where $\Omega_{o}$ is the angular velocity, $R_{o}$ the radius of the outer cylinder and $\nu$ the fluid’s kinematic viscosity) could be controlled to better than $\Delta Re= \pm 0.5$.

Unless perturbed the flow remains laminar up to Reynolds numbers exceeding $400$ confirming the concentricity and overall high accuracy of the set up. It is noted that in early prototypes of the experiment, misalignments of the cylinders had been found to trigger turbulence at significantly lower Re. In order to determine the critical point, where turbulence first becomes sustained, the laminar flow must be perturbed initially.  In our case this was achieved by a sudden increase to a higher Reynolds number $\Rey \approx 520$) followed by a quench to the lower target Reynolds number. Here the flow would settle to a state of spatio-temporal intermittency which was then left to stabilize for $10^7$ advective time units (see also Fig. S5) to approach its statistical steady state. Subsequently image time series were recorded for at minimum $10^7$ (closer to critical for up to $4 \cdot 10^7$) advective time units (see Fig.~\ref{fig:expSetUp}b). In accordance with the above suggested DP analogy and as confirmed experimentally for pipe flow \cite{mukund_hof_2018}, the temporal transients (blue data set in Fig. S5) are hence substantially larger than the splitting and decay times of single stripes (blue dashed line in Fig. S5). Towards the higher end of the investigated Reynolds number interval the measurement volume is densely packed with turbulent stripes (see Fig.~\ref{fig:visu}c, $\Rey=349$) and the stripe spacing of the order of $\lambda \approx 70h$ is the smallest spacing stripes can assume \cite{prigent_large-scale_2002,samanta_experimental_2011}. In the DP context $\lambda$ therefore corresponds to the size of a single site and is the natural length scale of the problem (see also Fig. S4 and the discussion in the SM). With decreasing Reynolds number the density of stripes and hence the turbulent fraction TF, which is the order parameter, decrease (Figures ~\ref{fig:visu}c to a). As shown in Fig.~\ref{fig:TfMu}a TF decreases continuously as the critical point is approached (TF is shown as a function of the reduced Reynolds number $\epsilon = (\Rey - \Rey_{c}) / \Rey_{c}$). 

More precisely TF is found to scale with a power law, i.e. $TF \sim \epsilon^{\beta}$. The best fit (red dashed line in Fig.~\ref{fig:TfMu}a) results in a value of $\Rey_{c}=330.0 \pm 0.5, \beta = 0.59 \pm0.03$. This experimentally obtained value of $\beta$ is in excellent agreement with the corresponding critical exponent of the directed percolation universality class in two spatial dimensions (2+1D DP), which is $\beta^{DP}=0.583$, (shown by the black line in Fig.~\ref{fig:TfMu}a). In addition to the power law scaling of TF, directed percolation is characterized by distinct scaling relations for the correlation length and correlation time in the vicinity of the critical point. Hence the corresponding two correlation exponents establish if the transition falls into this universality class. To test this we next investigated the sizes of the laminar gaps along the azimuthal direction $N(l_{\theta})$ in the vicinity of the critical point. As shown in Fig.~\ref{fig:TfMu}b the gap size distribution also follows a power law and the flow is hence scale invariant \cite{shi_scale_2013}. Finally in Fig.~\ref{fig:TfMu}c we plot the duration of the laminar gaps in time $N(t)$ and again a power law is obtained. The corresponding critical exponents for the spatial and temporal gap distributions (best fits are given by the red dashed lines) are found to be $\mu_{\bot}^{\theta}=1.22 \pm0.02$ and $\mu_{\parallel} =1.58 \pm0.03$, while the values for directed percolation (black lines in the respective figure panels) are $\mu^{DP}_{\bot}=1.204$ and $\mu^{DP}_{\parallel}=1.549$ and hence the values are in close agreement. 

While these three exponents provide sufficient evidence that the transition to turbulence belongs to the 2+1D DP universality class, additional exponents can be obtained in separate experiments and hence allow an independent confirmation. By considering the laminar gap size distributions at various distances from the critical point the corresponding characteristic length scale has been determined as a function of the control parameter (see Fig.~\ref{fig:nuCoeff} and supplemental materials). This procedure has been carried out separately for the gap spacings in the azimuthal and axial directions and for the gap durations (temporal gaps). In DP the corresponding correlation length and time diverge with characteristic power laws (black lines in Fig.~\ref{fig:nuCoeff}) as the critical point is approached. The experimental data for Couette flow (blue symbols in Fig.~\ref{fig:nuCoeff}) are in close agreement with the DP prediction.  The fitted power laws (red lines) diverge at $\Rey_{c}=330.0$ (the critical point determined from the measurements in Fig.~\ref{fig:TfMu}a). 

The transition to turbulence has been under scientific scrutiny for over a century and although directed percolation has been suggested as a possible solution to the problem three decades ago, direct evidence has only recently been obtained for simplified geometries. The canonical cases, pipe, Couette and channel flow appeared out of reach due to the enormous temporal and spatial scales that dominate the transition process. Resolving these scales in pipe and channel flows where turbulent structures move at the mean flow speed, would require experiments of lengths easily exceeding $10^3$ km. Couette flow on the other hand, can be realized in a circular geometry with periodic boundary conditions and hence appears a more feasible target. Nevertheless the required aspect ratios pose a formidable hurdle. In the present case the fluid had to be confined to a 240 micron thin sheet with the overall physical dimensions closely matching those of a page of A4 paper (~10pt). The velocity difference across the thin fluid film on the other hand had to be increased beyond 10 meters per second to trigger turbulence. The difficulty in measuring critical exponents is not specific to turbulence but more broadly applies to nonequilibrium systems. Although directed percolation is believed to be relevant to numerous problems in diverse areas of the natural sciences \cite{hinrichsen_non-equilibrium_2000}, experimental verification of this universality class \cite{takeuchi_directed_2007} has been a longstanding problem \cite{hinrichsen_possible_2000}. The transition to turbulence offers the first example that is of broad practical relevance.

\begin{acknowledgments}
We thank T.Menner, T.Asenov, P. Maier and the Miba machine shop of IST Austria for their valuable support in all technical aspects. We thank Marc Avila for comments on the manuscript. This work was supported by a grant from the Simons Foundation (662960, B.H.). We acknowledge the European Research Council under the European Union's Seventh Framework Programme (FP/2007-2013)/ERC Grant Agreement 306589 for financial support. K.A. acknowledges funding from the Central Research Development Fund of the University of Bremen, grant number ZF04B /2019/FB04 Avila Kerstin ("Independent Project for Postdocs"). L.K. was supported by the European Union’s Horizon 2020 research and innovation programme under the Marie Sk\l odowska-Curie grant agreement No. 754411. 
\end{acknowledgments}

\bibliographystyle{apsrev}
\bibliography{BibIst1}

\end{document}


\title{Supplemental Material for: \\The phase transition to turbulence in spatially extended shear flows}

\author{Lukasz Klotz}
\affiliation{ Institute of Science and Technology Austria, Am Campus 1, 3400 Klosterneuburg, Austria}
\affiliation{Institute of Aeronautics and Applied Mechanics, Warsaw University of Technology,  Nowowiejska 24, 00-665 Warsaw, Poland}
\author{Gr\'{e}goire Lemoult}%
 \affiliation{ CNRS, UMR 6294, Laboratoire Onde et Milieux Complexes (LOMC)53, Normandie Université, UniHavre, rue de Prony, Le Havre Cedex 76058, France}

\author{Kerstin Avila}%
\affiliation{ University of Bremen, Faculty of Production Engineering, Badgasteiner Strasse 1, 28359 Bremen, Germany }
\affiliation{Leibniz Institute for Materials Engineering IWT, Badgasteiner Strasse 3, 28359 Bremen, Germany}  
\author{Bj\"orn Hof}
\thanks{bhof@ist.ac.at}
\affiliation{ Institute of Science and Technology Austria, Am Campus 1, 3400 Klosterneuburg, Austria}
\email{bhof@ist.ac.at}
\date{\today}

\maketitle

\onecolumngrid

\begin{figure*}[!htp]
	\centering
	\includegraphics[trim={0 -0.6cm 0 -1cm},width= 0.99\textwidth]{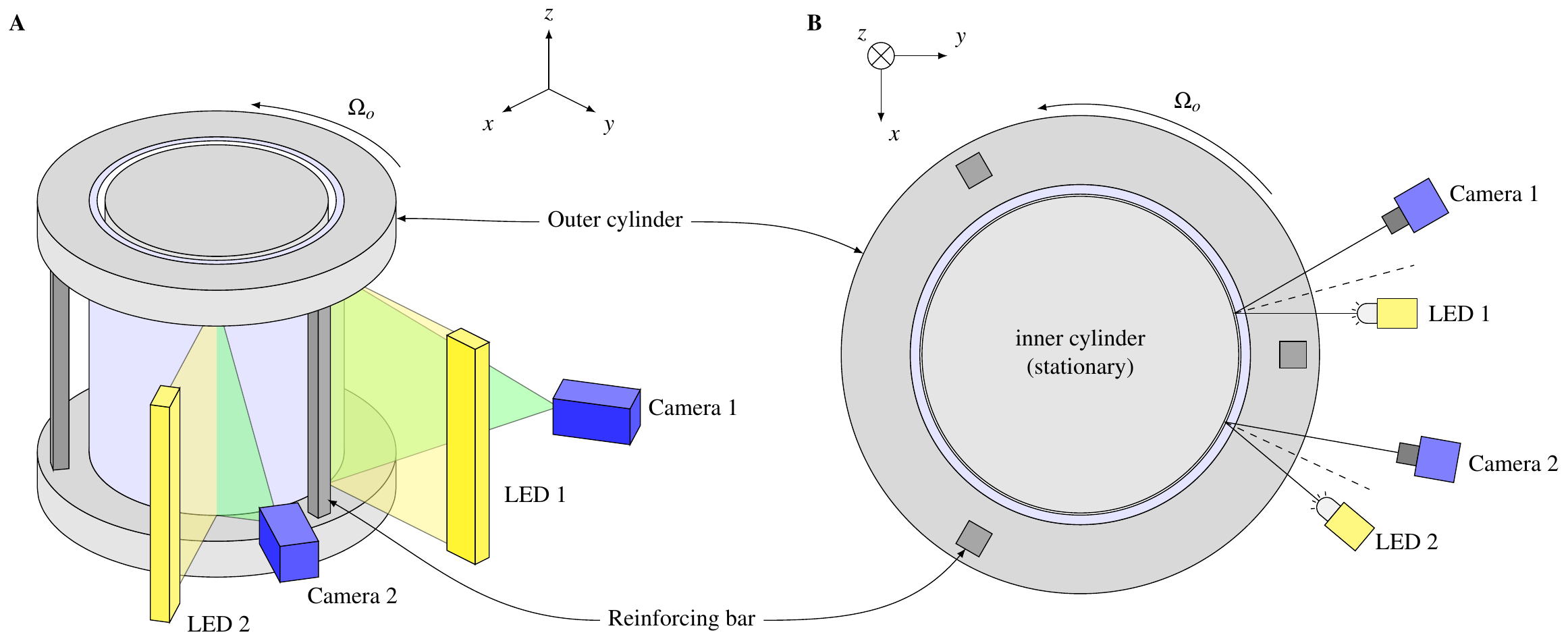}
	\caption{Schematic representation of the rotating outer cylinder with three reinforcing beams, along with the line cameras and line LED lamps. Viewed from the side (a) and from the top (b).}
	\label{fig:S1}
\end{figure*} 

\begin{figure*}[!htp]
	\centering
	\includegraphics[trim={0 -0.6cm 0 -1cm},width= 0.99\textwidth]{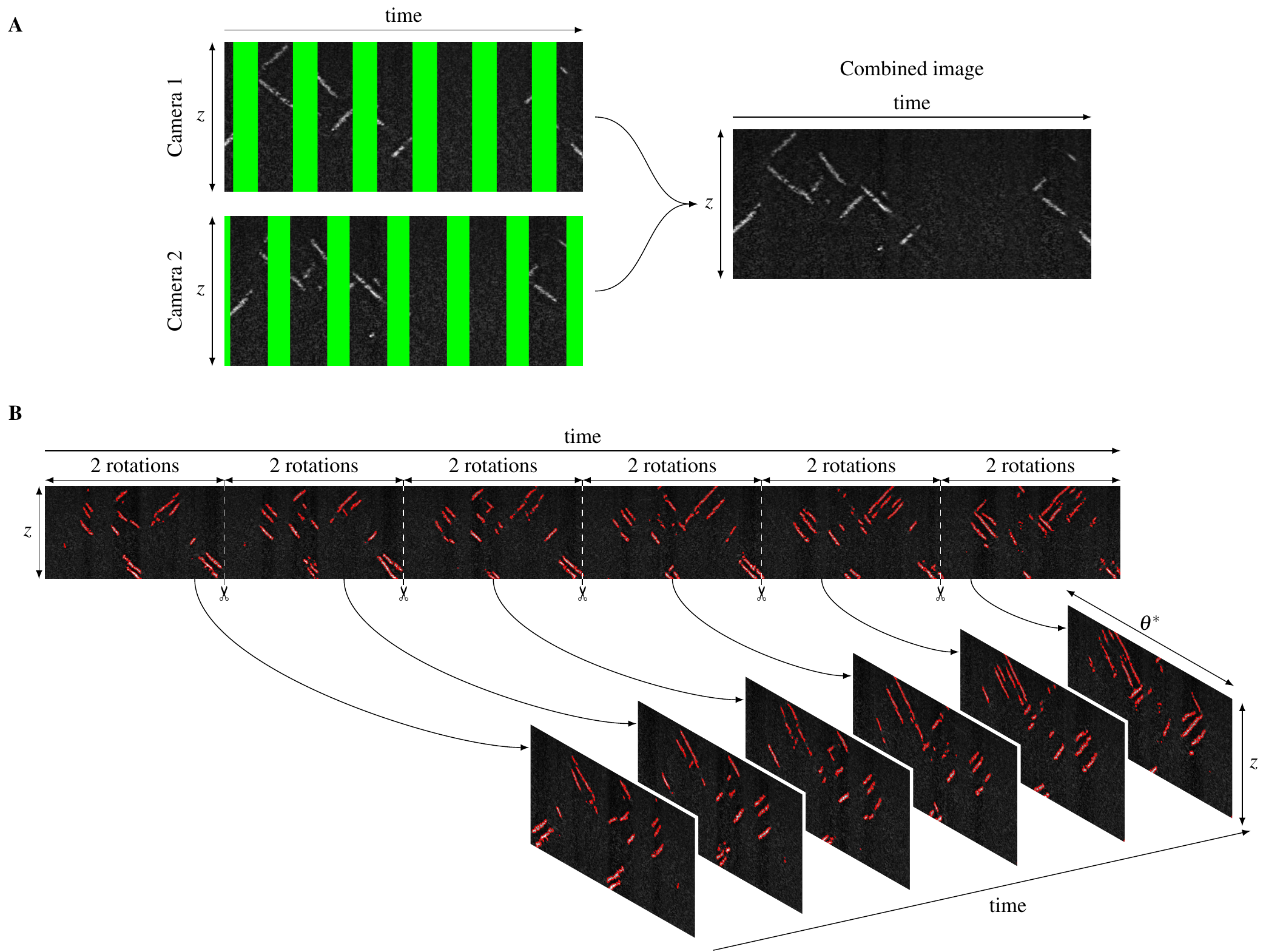}
	\caption{(a) Reconstruction of a single recording time step (corresponding to two rotations of the outer cylinder) from the images of the two line cameras. Green rectangles correspond to the areas obscured by the reinforcement beams; (b) illustration of the reconstruction of the spatio-temporal diagram and extraction of subsequent time steps. Flow field shown for $\Rey = 331$. }
	\label{fig:S2}
\end{figure*} 

\section{Experimental parameters}
The inner and outer cylinder radii were measured to be accurate to better than ±2.5µm and ±5µm, respectively. The gap uniformity during outer cylinder rotation (i.e. relative variation around the mean value) has been measured to be better than $\pm 6\mu$m. The eccentricity between inner and outer cylinder is less than $5\mu$m. The variation of the outer cylinder has been measured with a high precision leverage dial micrometer and the eccentricity between the inner and outer cylinders has been set using precision feeler gauge tape. Those are spatial errors and while they affect the system uniformity, they do not impact the average Reynolds number. Ethanol was used as the working fluid and the fluid’s kinematic viscosity $\nu$ was measured with an accuracy of $\pm 0.1\%$, in addition, from the calibration constant there is an uncertainty in the absolute value of $\pm 0.7\%$ . The temperature within the test section was controlled by circulating cooling fluid through the hollow inner cylinder and found to be constant and uniform to within $\pm 0.02$K. Resulting from above the uncertainty in the absolute value of the Reynolds number is ($\Rey = \Omega_oR_oh / 2\nu$) is $\pm 14$. The relative error of the average Reynolds number, i.e. the control parameter, is far smaller and only amounts to $\pm 0.5$.

\section{Flow visualization}

To measure the different exponents associated with DP, the flow pattern has to be monitored and resolved in space and time. The turbulent flow structures can be visualized and discriminated from the laminar background flow by seeding the fluid with a small amount of anisotropic reflective flakes. The patterns can then be readily recorded by ordinary high speed cameras. However in practice the cylindrical geometry makes the instantaneous capture of the full flow pattern more difficult. We circumvent this difficulty by taking advantage of the large time scale separation between the characteristic advection time of turbulence and the characteristic time associated with the dynamical changes of the turbulent pattern. Turbulent patches are advected at the mean flow speed  , and will return to the same azimuthal position, after a time , corresponding to 2 outer cylinder rotations. Expansion and growth of stripes are slow compared to this time scales and in particular decay and splitting times are orders of magnitude longer. Hence the turbulent flow pattern can be regarded as close to frozen over a full revolution (Taylor hypothesis). See also \cite{lemoult_directed_2016} for more details. Therefore, the full azimuthal flow pattern can in principle be reconstructed from a stationary (in the laboratory frame of reference) line camera (axially aligned) run at sufficiently high frequency.

Due to the high rotation speeds of the outer cylinder the experimental set up had to be reinforced by three steel bars connecting the top to the bottom end plates  (see Fig.~S \ref{fig:S1}) and co-rotating with the outer cylinder. These bars partially obscured the camera view. Therefore, a second line scan camera was implemented at a different location viewing the flow from an appropriate angle to capture those regions obscured in camera 1. The full spatio-temporal flow field can then be reconstructed by combining the information from both cameras (Fig.~S \ref{fig:S2}). The validity of this reconstruction method and more generally the applicability of the Taylor hypothesis applied is apparent from the continuous evolution of the stripe patterns (e.g. see movie S4).

\section{Characteristic length/time}

\begin{figure*}[!htp]
	\centering
	\includegraphics[trim={0 -0.6cm 0 -1cm},width= 0.99\textwidth]{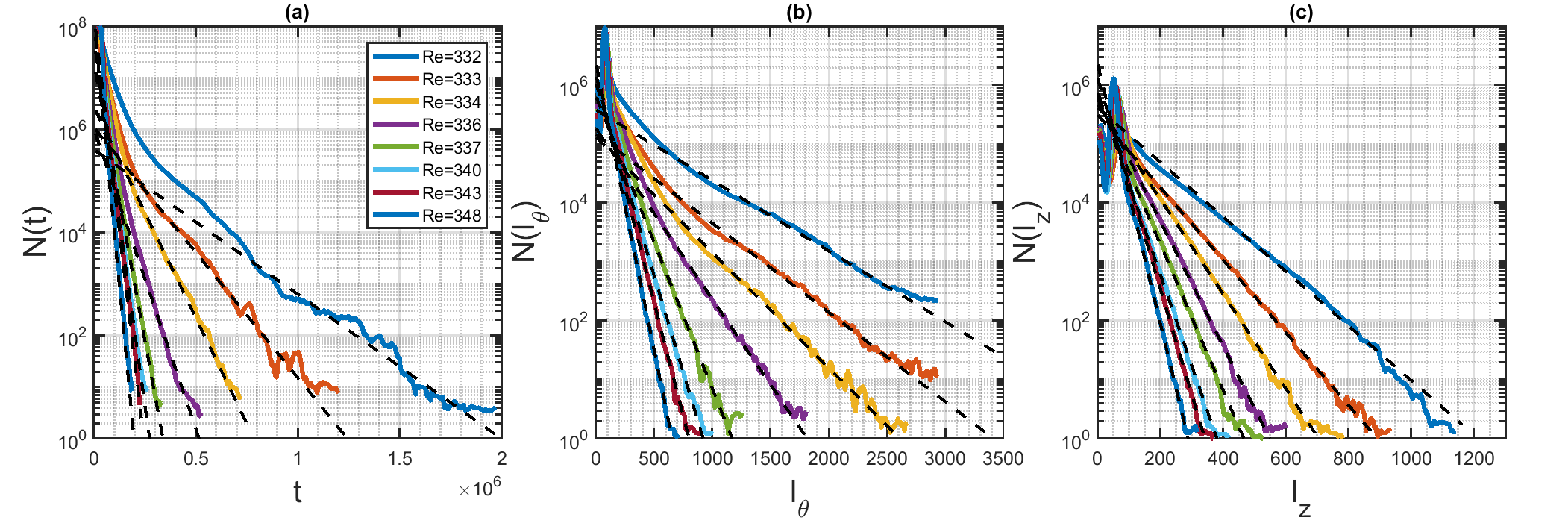}
	\caption{ Distribution of the laminar time intervals $N(\tau)$ (panel a), distribution of the sizes of laminar gaps along the azimuthal (panel b: $N(\l_{\theta})$) and axial (panel c: $N(\l_{z})$) directions. Exponential tails are marked by black dashed lines, obtained using the best fit to the function in the form: a) $N(\tau)\sim exp(-\tau/\xi_{\parallel})$, b) $N(\l_{\theta}) \sim exp(-\l_{\theta}/\xi_{\bot})$, and c) $N(\l_{z}) \sim exp(-\l_{z}/\xi_{\bot})$, respectively. $\xi_{\parallel}$ and $\xi_{\bot}$ indicates correlation time and correlation length.}
	\label{fig:S3}
\end{figure*} 

Close to the critical point the distributions of laminar gaps become scale invariant and hence follow universal power laws. Away from the critical point the corresponding size distributions fall off exponentially. The variation of these exponential tails with distance to the critical point can be used to estimate the spatial and temporal correlation exponent as shown in the following. The corresponding distributions of the laminar gap sizes in the azimuthal direction, $N(\l_{\theta})$, in the axial direction, $N(\l_{z})$, and in time, $N(\tau)$, are plotted in Fig. 3 for various Reynolds numbers. The correlation time ($\xi_{\parallel}(\Rey)$) and correlation lengths ($\xi_{\bot}^{\theta}(\Rey)$ and $\xi_{\bot}^{z}(\Rey)$) correspond to the slope of the respective exponential tail (dashed lines in Fig.~S \ref{fig:S3}a,b,c). Finally the variation of the correlation lengths allows to determine the correlation exponents as shown in Fig. 4.

\begin{figure*}[!htp]
	\centering
	\includegraphics[trim={0 -0.6cm 0 -1cm},width= 0.69\textwidth]{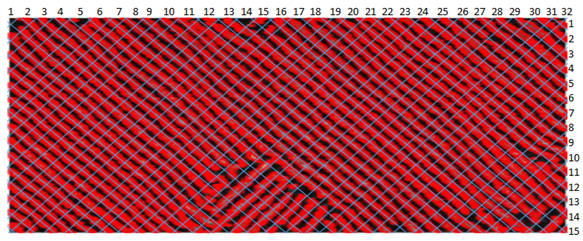}
	\caption{ Instantaneous snapshot of a test section that corresponds to fully occupied domain in Couette flow.}
	\label{fig:S4}
\end{figure*} 

\section{Spatial requirements}

According to the recently proposed analogy between transitional turbulence and directed percolation an active site in the DP sense corresponds to the localized structure turbulence naturally assumes in the transitional regime, a puff in spatially one dimensional flows and a stripe in two dimensions. More precisely it is the minimal distance between two such localized turbulent structures, i.e. between two puffs or two stripes that sets the grid spacing. Fig.~S \ref{fig:S4} shows one realization of a fully occupied domain for Couette flow. Note that in two dimensional flows the orientation of stripes does not matter for this analogy and that the fully occupied state is not a unique pattern. Any dense stripe pattern corresponds to a fully occupied lattice in a sense that no new turbulent stripe (or segment thereof) can be created unless an existing stripe decays and makes space for a new one to arise. Given that the minimum stripe spacing corresponds to approximately 70 h even a domain of 1000 h x 1000 h which would be considered large in the context of traditional transition studies only consists of 10 x 10 sites in the DP sense.  

\section{Temporal requirements}

\begin{figure*}[!htp]
	\centering
	\includegraphics[trim={0 -0.6cm 0 -1cm},width= 0.99\textwidth]{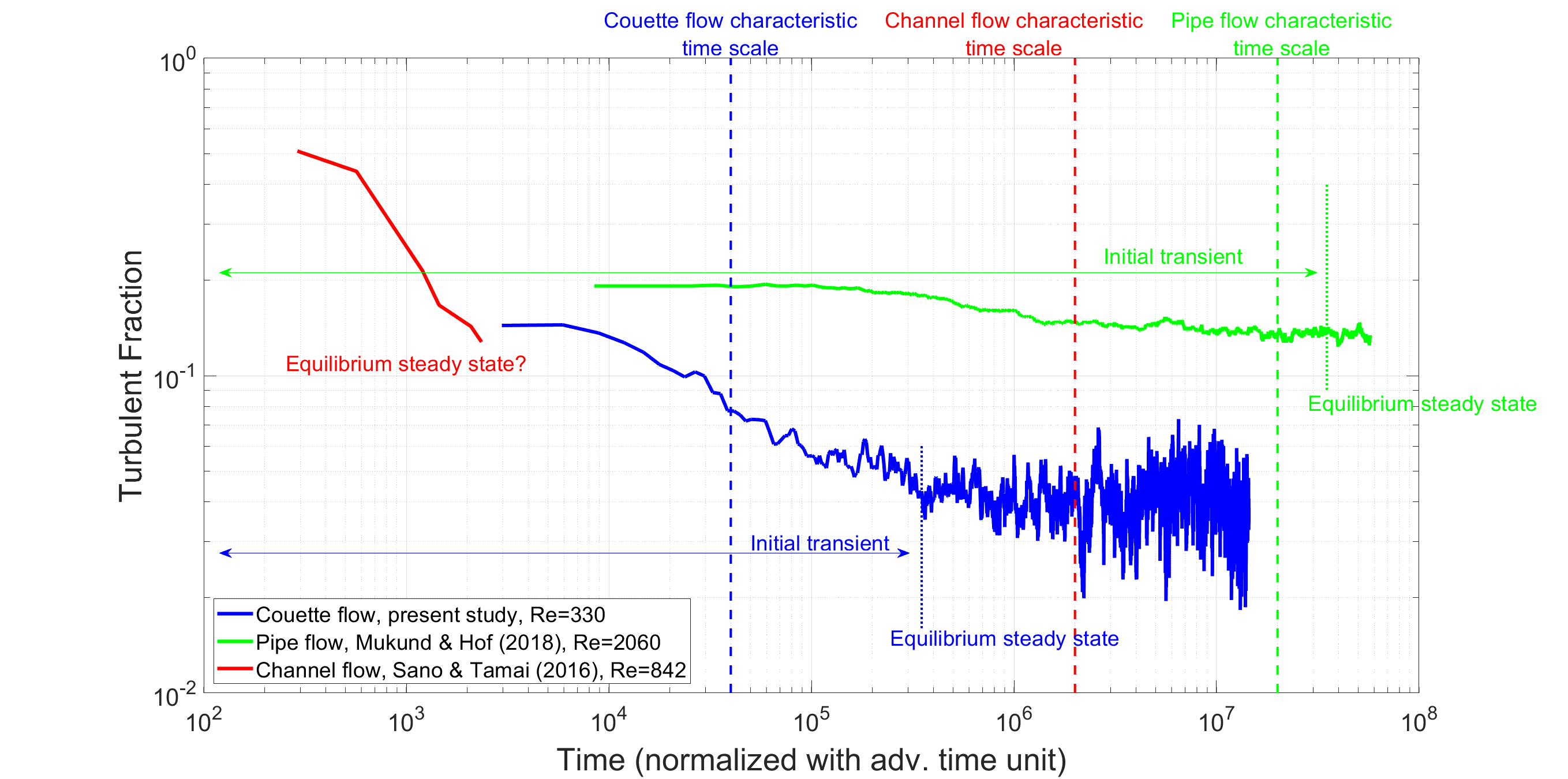}
	\caption{ Illustration of evolution of the turbulent fraction for Couette (blue), channel (red, inferred from \citep{sano_universal_2016}), and pipe (green, inferred from \citep{mukund_hof_2018}) flow. Characteristic time scales that correspond to one 'DP time step' are marked by vertical dashed lines in corresponding colors. Vertical dotted lines indicate the initial transient evolution observed experimentally in Couette (blue dotted line) and in pipe (green dotted line) flows}
	\label{fig:S5}
\end{figure*} 

The dynamics of stripe patterns are to a first approximation governed by the characteristic time scales of the decay and the proliferation of its individual components, i.e. in the present case decay and splitting of individual stripes. In particular in the vicinity of the critical point these time scales tend to be very large but vary between flows. In pipe flow decay and splitting times are of order $10^7$ advective time units, in channel flow $10^6$ whereas in Couette flow they happen to be considerably shorter ( of order $10^4$ advective time units). In the DP analogy this time scale then is proportional to a single time step and consequently any study to test this analogy must resolve time scales far larger than those characteristic scales of individual stripes (/puffs). Fig.~S \ref{fig:S5} shows that average quantities of stripe patterns such as in this case the turbulent fraction evolves on excessively large time scales that as would be expected correspond to multiple ‘DP time steps’. In pipes and channel flows a central limitation is that turbulence is advected downstream at the mean flow speed. This essentially makes it in practice impossible to follow individual turbulent structures for the time scales required. For example in case of the recent channel experiments of \citep{sano_universal_2016} the experimental set up was 2500 h long (streamwise). Given that turbulent structures move by one h in one advective time unit the maximum time that a turbulent structure could be observed in their case corresponded to 2500 advective time units which is three orders of magnitude shorter than the ‘DP time step’ for channel flow.




Movie S4.
Illustration of the dynamics of the turbulent fraction recorded for $\Rey = 331$.

\begin{acknowledgments}
We thank T.Menner, T.Asenov, P. Maier and the Miba machine shop of IST Austria for their valuable support in all technical aspects. We thank Marc Avila for comments on the manuscript. This work was supported by a grant from the Simons Foundation (662960, B.H.). We acknowledge the European Research Council under the European Union's Seventh Framework Programme (FP/2007-2013)/ERC Grant Agreement 306589 for financial support. K.A. acknowledges funding from the Central Research Development Fund of the University of Bremen, grant number ZF04B /2019/FB04 Avila Kerstin ("Independent Project for Postdocs"). L.K. was supported by the European Union’s Horizon 2020 research and innovation programme under the Marie Sk\l odowska-Curie grant agreement No. 754411. 
\end{acknowledgments}

\bibliographystyle{apsrev}
\bibliography{BibIst1}